\documentclass[11pt, oneside, onecolumn]{IEEEtran}
\IEEEoverridecommandlockouts

\usepackage[pdftex]{graphicx}
\usepackage{dsfont}  		

\usepackage[cmex10]{amsmath}

\usepackage{anysize}

\usepackage{algpseudocode}

\usepackage[tight,footnotesize]{subfigure}

\usepackage[caption=false,font=footnotesize]{subfig}
\usepackage{epstopdf}
\usepackage{url}

\marginsize{1.6cm}{1.6cm}{0.4cm}{1.6cm} 

\usepackage[latin1]{inputenc}

\usepackage{amsfonts}
\usepackage{amssymb}
\usepackage{mathrsfs}

\begin{document}

\title{\vspace{0mm}A Stochastic Approach for Resource Allocation with Backhaul and Energy Harvesting Constraints}

\author{\IEEEauthorblockN{Javier Rubio, Olga Mu\~noz, and Antonio Pascual-Iserte}
\IEEEauthorblockA{\\Department of Signal Theory and Communications \\Universitat Polit\`ecnica de Catalunya (UPC), Barcelona, Spain\\ Emails:\{javier.rubio.lopez, olga.munoz, antonio.pascual\}@upc.edu}
\thanks{The research leading to these results has received funding from the European Commission in the framework of the FP7 Network of Excellence in Wireless COMmunications NEWCOM\# (Grant agreement no. 318306) and the project TUCAN3G (Grant agreement no. ICT-2011-601102), from the Spanish Ministry of Economy and Competitiveness (Ministerio de Econom\'ia y Competitividad) through the project TEC2011-29006-C03-02 (GRE3N-LINK-MAC), project TEC2013-41315-R (DISNET), and FPI grant BES-2012-052850, and from the Catalan Government (AGAUR) through the grant 2014 SGR 60.}\vspace{-7mm}}

%

\maketitle

\vspace{-5mm}

\begin{abstract}
We propose a novel stochastic radio resource allocation strategy that achieves long-term fairness considering backhaul and air-interface capacity limitations. The base station is considered to be only powered with a finite battery that is recharged by an energy harvesting source. Such energy harvesting is also taken into account in the proposed resource allocation strategy. This technical scenario can be found in remote rural areas where the backhaul connection is very limited and the base stations are fed with solar panels of reduced size. Our results show that the proposed scheme achieves higher fairness among the users and, in some cases, a higher sum-rate compared with the well-known proportional fair scheduler.

\end{abstract}

\vspace{-5mm}
\section{Introduction}
We consider in this paper a downlink (DL) radio resource allocation strategy for a system with limited backhaul capacity and where the base station (BS) is equipped with a finite battery recharged by an energy harvesting source. Although backhaul availability has been taken for granted in conventional systems, backhaul is, in general, a limited resource. This is the case, for instance, of the deployment planned in the European project TUCAN3G project (http://www.ict-tucan3g.eu) in which 3G femtocells empowered by solar panels of limited size and connected to the core network through a limited capacity WiFi-LD backhaul are used to provide 3G connectivity to users located in rural remote locations in Per\'u. Such limited WiFi-LD connections already exist currently and are used basically to provide remote health services. The final \emph{social} objective of the project is to contribute to the economical development of such areas through the provision of communication services to the general users beyond the current limited health services.
  
The backhaul capacity limitation can be introduced in the resource allocation problem by imposing a maximum instantaneous aggregate traffic rate constraint \cite{chowdhery:11}, \cite{zhou:13}, \cite{zcui:11}. However, in real deployments, the backhaul capacity can only be measured in average terms. In addition, limiting the sum-rate instantaneously at each specific scheduling period may hamper the performance of the system in terms of the achievable long-term rates. In these circumstances, it seems less limiting to use high data rates in the access network whenever the channel conditions allow (even using greater instantaneous values than the average constraint imposed by the backhaul) provided that the average backhaul rate constraint is met when averaging the traffic served in several scheduling periods. Note that the backhaul constraint in terms of average traffic is suitable if we assume that queues are implemented at the entrance of the access network.

In addition to the backhaul limitation, the energy available at the BS may be a very limited resource as well. If the BS is only powered with batteries (as may happen in rural environments, for instance in the TUCAN3G deployments described above), then the battery status as well as the harvesting capabilities (if any) should also be considered in the scheduling strategy explicitly if we want to optimize the performance subject to the energy limitations. In the scenario that we will consider in this paper, the BS will be powered only by a limited battery and an energy harvesting device, e.g., solar panels, which will allow recharging the battery \cite{paradiso:05}.

In this paper we propose a long term fairness scheduler that considers a long-term backhaul constraint, the battery status of the BS, and the energy that it is being harvested. When there is no reason for treating flexible service rate users differently, the maximin criterion is a meaningful scheduling approach \cite{rhee:00}. This approach maximizes, at each scheduling period, the minimum of the throughputs of the users. Essentially our goal is to provide a balanced long-term rate to a set of users. In addition, and as commented previously, the scheduler will consider explicitly the energy and backhaul constraints, while taking advantage in an opportunistic way of the instantaneous good wireless channel conditions.


The remainder of this paper is organized as follows. In Section \ref{sec_sysm} we describe the system model. Section \ref{sec_pro} presents the resource allocation strategy developed in the paper. The numerical evaluation, presented in Section \ref{sec_num}, has been carried out using realistic models taken from remote rural locations in the forest in Per\'u. Finally, conclusions are drawn in Section \ref{sec_con}.

\vspace{-7mm}
\section{System Model}
\label{sec_sysm}
\subsection{System Description}
Let us consider a DL scenario composed of a single BS and several users. We assume that the system is based on WCDMA technology \cite{goldsmith} and two different types of users coexist: voice users and data users. Let us denote the set of voice and data users by $\mathcal{K}_V$ and $\mathcal{K}_D$, respectively. We assume that voice users request a fixed service rate whereas data users request a flexible service rate. 

Users in WCDMA are multiplexed using codes \cite{goldsmith}. We assume that the network operator has already reserved a set of codes for the voice users and the remaining codes are to be allocated among the data users. Thus, the amount of available codes in each set is known and fixed at the BS. 

The BS is solely powered with a battery and an energy harvesting source. The energy harvesting source allows the BS to collect energy from the environment and recharge the battery (for example, solar panels). This is especially important in rural areas, where the access to the power grid may be impossible or too expensive. We consider that only causal information is available for the resource allocation strategy, i.e., only information of the past and current harvesting collections and battery dynamics will be available to execute the scheduling strategy at each particular scheduling period, yielding to an \emph{online approach}.

\begin{figure}[t]
\centering
\includegraphics[width = 0.6\textwidth]{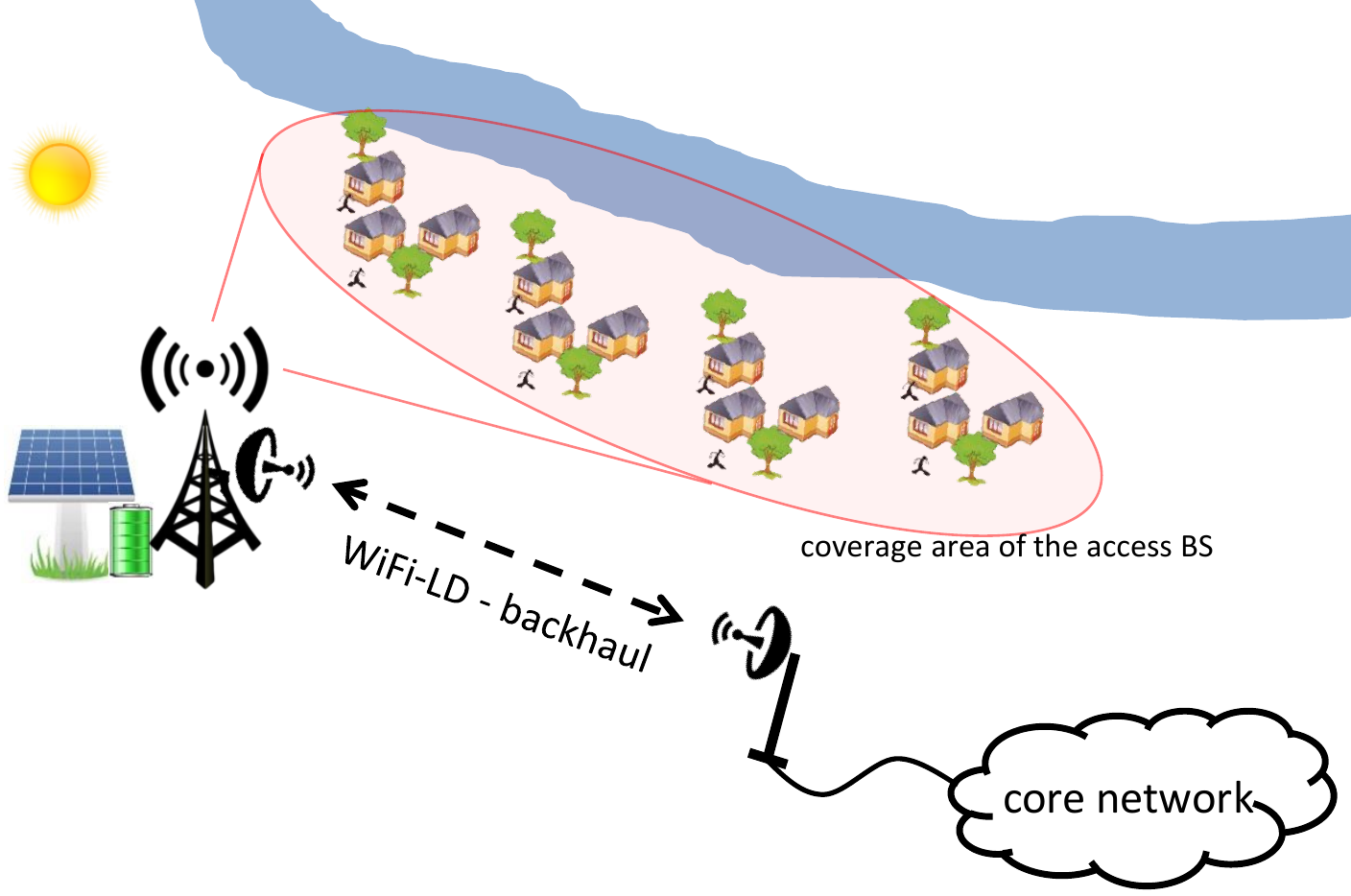}
\vspace{-2mm}
\caption{Architecture of the target rural scenario under consideration in the paper. The BS is powered with a solar panel and a battery and the backhaul considered is based on WiFi-LD. The specific details of the real deployment as well as the location will be explained in the simulation section.}
\label{arch}
\vspace{-3mm}
\end{figure}

One of the novelties of this work is that we account for a maximum backhaul rate constraint. However, instead of limiting the instantaneous access network data rates as the maximum flow allowed by the backhaul, as in \cite{chowdhery:11}, \cite{zhou:13}, and \cite{zcui:11}, we limit the average throughput served by the access network. That means that we allow the instantaneous rate in the access wireless links to surpass the backhaul limitation. This can be done whenever we have queues at the entrance of the access network and such queues are stable (which, in fact, is guaranteed by imposing that the average aggregated rate is not higher than the backhaul capacity). Moreover, as it usually happens with backhaul connections in real deployments, it is not possible in the transport layer to differentiate among different users with the same type of service, for example in DiffServ-based networks \cite{shioda:05}. Accordingly, we consider in the resource allocation problem in the access network that the backhaul capacity is equally divided among the users with the same type of service. Fig. \ref{arch} presents the system architecture of the target rural scenario.

\vspace{-3mm}
\subsection{Power Consumption Model and Battery Dynamics}
In this subsection, we introduce the power consumption and the battery model considered in this paper. The overall power consumption at the BS is modeled as the addition of the radiated power, which is divided into the power devoted to pilot channels $(P_{\text{CPICH}}$ assumed to be fixed$)$ and the power consumed by the traffic channels $(P_{\text{BS}}(t))$, and a fixed power consumed by the electronics of the BS $(P_{\text{c}})$, where $t$ denotes the scheduling period. The model considered for the last term is based on \cite{auer:11} and includes the power consumption of the radio frequency (RF) chains, the baseband power consumption, and the consumption of the cooling systems. The maximum traffic power will be forced to depend on the current battery level of the BS, as it will be apparent later.

Given that, the overall energy consumption by the BS during the $t$-th scheduling period is
\begin{equation}
E(t) \triangleq T_s \cdot \left(P_{\text{CPICH}}  + P_{\text{BS}}(t) +  P_{\text{c}}\right), \quad \forall t,
\label{energy}
\end{equation}
where $T_s$ is the duration of the scheduling period. We consider that the amount of power that can be used for traffic services is limited (due to physical constraints of the amplifiers of the BS equipment). Such maximum traffic power will be denoted as $P^{\max}_{\text{BS}}$, so $P_{\text{BS}}(t)\le P^{\max}_{\text{BS}}$. 

Let $B(t)$ be the energy stored at the battery of the BS at the beginning of the scheduling period $t$. Then at period $t+1$, the battery level is updated in general as
\begin{equation}
B(t+1) = f(B(t), E(t), H(t)), \quad \forall t,
\end{equation}
where $H(t)$ is the energy harvested in Joules during the scheduling period $t$ and the function $f(\cdot): \mathbb{R}_+ \times \mathbb{R}_+\times \mathbb{R}_+ \rightarrow \mathbb{R}_+$ depends upon the battery dynamics, such as storage efficiency and memory effects. A common practice is to consider the following battery update:
\begin{equation}
B(t+1) =  \left(B(t) - E(t) + H(t)\right)^{B^{\max}}_0, \quad \forall t,
\label{bat_eq}
\end{equation}
where $(x)^b_a$ is the projection of $x$ onto the interval $[a, b]$, i.e., $(x)^b_a = \min\{\max\{a,x\},b\}$, which accounts for possible battery overflows and assures that the battery levels are non-negative, and $B^{\max}$ is the battery capacity. Notice that the whole harvesting collected during period $t$ is assumed to be available in the battery at the end of the period for simplicity. In general, the total energy consumed by the BS during one period will be limited by a function of the current battery level as
\begin{equation}
T_s \cdot \left(P_{\text{CPICH}}  + P_{\text{BS}}(t) +  P_{\text{c}}\right) \le g(B(t)), \quad \forall t,
\label{fung}
\end{equation}
where the function $g(\cdot)$ is defined as $g(B(t)) \triangleq \min\{T_s \left(P_{\text{CPICH}}  + P^{\max}_{\text{BS}} +  P_{\text{c}}\right), w(B(t))\}$, and $w(\cdot) : \mathbb{R}_+ \rightarrow \mathbb{R}_+$ a generic continuous increasing function that fulfills that $w(B(t)) \le B(t),$ $\forall t$. For example, if all the battery is allowed to be spent during one particular epoch, then $w(B(t)) = B(t)$. Nevertheless, the approach followed in this paper is to limit the amount the battery that can be used in one particular period in order to use the energy in a more conservative way. Thus, we consider that only a given fraction of the battery is allowed to be used in a particular scheduling period, i.e.,
\begin{equation}
w(B(t)) = \alpha\cdot B(t), \quad 0 \le \alpha \le 1.
\end{equation}

\vspace{-10mm}
\subsection{Energy Harvesting Model}
We assume a discretized model for the energy arrivals \cite{yang:12} where $H(t)$ is modeled as an ergodic Bernoulli process (which is a particular case of a Markov chain). As a result, only two values of harvested energy are possible, i.e., $H(t) \in \{0,e\}$, where $e$ is the amount of Joules contained in an energy packet. The probability of receiving an energy harvesting packet during one scheduling period depends on the actual harvesting intensity (in the case of solar energy, it depends on the particular hour of the day) and is denoted by $p(t)$. Note that a higher value of $p(t)$ will be obtained in scheduling periods where the harvesting intensity is higher, e.g., during solar presence such as during the day, and a lower value of $p(t)$ will be obtained during periods of solar absence, such as during the night. 

\vspace{-3mm}
\subsection{System Assumptions}
Let us collect all the channel gains, $h_k$ that includes the antenna gains, the path loss, and the fading, in $\textbf{h} = \{h_k, \,\forall k\in\mathcal{K}_V\cup\mathcal{K}_D\}$. Generally, the wireless channels depend on the specific scheduling period, $\textbf{h}(t)$, as they vary over time but for simplicity in the notation, we will just refer to them as $\textbf{h}$ throughout the paper. The traffic power, $P_{\text{BS}}(t)$ from \eqref{energy}, can be split into power for voice and data connections as $P_{\text{BS}}(t) = \sum_{k\in\mathcal{K}_V}\check{p}_k(\textbf{h}) + \sum_{k\in\mathcal{K}_D} p_k(\textbf{h})$, where $\check{p}_j(\textbf{h})$ and $p_k(\textbf{h})$ are the instantaneous powers corresponding to the transmission toward the $j$-th and $k$-th voice and data user, respectively. Let $P_{\text{RAD}}(t) = P_{\text{BS}}(t) + P_{\text{CPICH}}$ be the overall radiated power by the BS.  

The set of voice users request a fixed data rate and we assume that just one WCDMA code is assigned to them. This is translated into a minimum signal to interference and noise ratio (SINR) requirement as follows:
\begin{equation}
\frac{M_V\check{p}_k(\textbf{h}))h_k}{\theta (P_{\text{RAD}}(t) - \check{p}_k(\textbf{h}))h_k + \sigma^2} \ge \Gamma_k, \quad \forall k \in \mathcal{K}_V,
\end{equation}
where $M_V$ is the spreading factor for voice codes, $\theta$ is the orthogonality factor among DL codes \cite{goldsmith}, and $\sigma^2$ is the noise power. For simplicity in the notation and tractability, we will consider the following approximation throughout the paper:\footnote{If the number of users is relatively high,  $P_{\text{RAD}}(t) \gg \check{p}_k$, and the approximation is fair. In any case, the approximation provides a lower bound of the actual SINR value.}
\begin{equation}
\theta(P_{\text{RAD}}(t) - \check{p}_k(\textbf{h}))h_k + \sigma^2 \approx \theta P_{\text{RAD}}(t)h_k + \sigma^2. 
\label{approx}
\end{equation}

On the other hand, the set of data users request a flexible service rate. The instantaneous throughput in the wireless access channel achieved during one particular scheduling period by the $k$-th user, $r_k(\textbf{h})$, is upper bounded by the maximum achievable rate that the access network is able to provide, which is formulated as
\begin{equation}
r_k(\textbf{h}) \le n_k(\textbf{h}) \frac{W}{M_D}\log_2\left(1+\frac{M_Dp_k(\textbf{h})h_k}{n_k(\textbf{h})(\theta P_{\text{RAD}}(t) h_k + \sigma^2)}\right),
\end{equation}
where $M_D$ is the spreading factor for data codes, $W$ is the chip rate, and $n_k(\textbf{h})$ is the number of codes assigned to user $k$. Notice that we have also approximated the denominator within the logarithm as in \eqref{approx}.

\section{Problem Formulation}
\label{sec_pro}

Let us introduce the following set of definitions: $\textbf{r} \triangleq \{r_k(\textbf{h}),\,\, \forall k\in \mathcal{K}_D\}$, $\check{\textbf{p}} \triangleq \{\check{p}_k(\textbf{h}),\,\, \forall k\in \mathcal{K}_V\}$, $\textbf{p} \triangleq \{p_k(\textbf{h}),\,\, \forall k\in \mathcal{K}_D\}$, $\textbf{n} \triangleq \{n_k,(\textbf{h})\,\, \forall k\in \mathcal{K}_D\}$. We formulate an optimization problem for the resource allocation strategy with backhaul and energy constraints to be executed at the beginning of each particular scheduling period, which involves finding the optimum resource allocation variables, $\textbf{r}$, $\check{\textbf{p}}$, $\textbf{p}$, and $\textbf{n}$ that maximize the minimum of the expected throughputs (note that if a scheduling criterion different from the maximin approach is to be taken, problem \eqref{pro_1} could be extended by just reformulating the objective function accordingly):
\begin{alignat}{3}
\mathop{\text{maximize}}_{\textbf{r},\, \check{\textbf{p}},\, \textbf{p},\, \textbf{n},\, P_{\text{RAD}}(t)}& \quad \min_{k\in \mathcal{K}_D}\,\, \mathbb{E}_{\textbf{h}}[r_k(\textbf{h})] \label{pro_1}\\
\textrm{subject to}
& \quad C1: \frac{M_V\check{p}_k(\textbf{h})h_k}{\theta P_{\text{RAD}}(t) h_k + \sigma^2} \ge \Gamma, \quad \forall k \in \mathcal{K}_V\nonumber\\
& \quad C2:\mathbb{E}_{\textbf{h}}[r_k(\textbf{h})] \le \frac{R_{BH}-\check{R}_{BH}(|\mathcal{K}_V|)}{\xi|\mathcal{K}_D|}, \quad \forall k \in \mathcal{K}_D\nonumber \\
& \quad C3:r_k(\textbf{h}) \le n_k(\textbf{h})  \frac{W}{M_D}\log_2\left(1+\frac{M_Dp_k(\textbf{h})h_k}{n_k(\textbf{h})(\theta P_{\text{RAD}}(t) h_k + \sigma^2)}\right),\quad  \forall k \in\mathcal{K}_D\nonumber \\
& \quad C4: T_s\left(\sum_{k\in\mathcal{K}_V}\check{p}_k(\textbf{h}) + \sum_{k\in\mathcal{K}_D} p_k(\textbf{h}) \right) \le \phi \left(B(t)\right) \nonumber \\
& \quad C5: \sum_{k\in\mathcal{K}_D} n_k(\textbf{h}) \le N_{\max} \nonumber \\
& \quad C6: r_k(\textbf{h}) \ge 0, \,p_k(\textbf{h}) \ge 0, \,n_k(\textbf{h}) \ge 0,\quad \forall k \in \mathcal{K}_D \nonumber \\
& \quad C7: P_{\text{RAD}}(t) = \sum_{k\in\mathcal{K}_V}\check{p}_k(\textbf{h}) + \sum_{k\in\mathcal{K}_D} p_k(\textbf{h}) + P_{\text{CPICH}}, \nonumber
\end{alignat}
where $\xi$, $(\xi>1)$, is an overhead considered for the data transmissions to be sent through the backhaul, $\check{R}_{BH}(|\mathcal{K}_V|)$ is the backhaul capacity used by the voice users\footnote{The overall backhaul capacity required for a set of voice users to be sent through the backhaul generally depends on the current number of voice users being served. In real deployments, voice users can be jointly encoded and, thus, the overall overhead for voice users may be reduced as the number of voice users increases. Anyway, in the problem formulation and the for the sake of generality, we just use the notation $\check{R}_{BH}(|\mathcal{K}_V|)$.}, being $|\mathcal{K}_V|$ the number of voice users, $R_{BH}$ is the overall backhaul capacity, $\Gamma$ is the target SINR for the voice users, the function $\phi(\cdot)$ is related to $g(\cdot)$ in \eqref{fung} as $\phi \left(B(t)\right) = g(B(t)) - T_s\cdot \left( P_{\text{CPICH}} + P_c\right)$, and $N_{\max}$ is the number of available codes for the data users. Although all the variables in the optimization problem \eqref{pro_1} depend on the scheduling period $t$, we only keep such explicit dependence w.r.t. time in variable $P_{\text{RAD}}(t)$ to make explicit that that temporal evolution of the battery levels has a direct impact on the maximum power to be spent for the voice and data traffic, which is not constant along time.

It is important to realize that problem \eqref{pro_1} may not be feasible due to constraint $C1$ as it may happen that there could not be enough power to satisfy all the target SINR simultaneously. However, let us consider initially through the development that the problem is feasible (the feasibility condition will be developed later on the paper). Notice that, at the optimum, $C4$ is attained with equality. Otherwise, if $C4$ is not fulfilled with equality, we could re-scale all the powers with a common positive factor higher than 1 until $C4$ is fulfilled with equality. This would increase the objective function and all the other constraints would still be fulfilled. Because of this reason, we can assume that the optimum value of $P_{\text{RAD}}(t)$ is $P^\star_{\text{RAD}}(t) = \frac{\phi \left(B(t)\right)}{T_s} + P_{\text{CPICH}}$ and we can eliminate constraint $C7$ from problem \eqref{pro_1}. Constraint $C2$ states that the average throughput that a user is experiencing in the access network should not exceed the maximum backhaul rate assigned to such user (every user has been already assigned a portion of the backhaul, as commented before). Again we state that this is so because, in general, backhauls are not able to differentiate among users. If the backhaul was able to do so, then $C2$ could be rewritten as $\sum_{k\in\mathcal{K}_D}\mathbb{E}_{\textbf{h}}[r_k(\textbf{h})] \le \frac{R_{BH}- \check{R}_{BH}(|\mathcal{K}_V|)}{\xi} $. In any case, notice that the instantaneous rates allocated to one user in the access network can be higher in some scheduling periods than the maximum backhaul per-user rate $\left(\frac{R_{BH}- \check{R}_{BH}(|\mathcal{K}_V|)}{\xi|\mathcal{K}_D|}\right)$ thanks to the fact that queues are considered at the entrance of the access network. The average rate constraint $C2$ assures that the queues will be stable.

It is easy to realize that the problem is separable into voice and data users without loss of optimality. For this reason, we start by analyzing the voice users.

\subsection{Resource Allocation for Voice Users}
Voice users must satisfy a minimum SINR constraint that is related to the target data rate service. Such constraints are:
\begin{equation}
\frac{M_V\check{p}_k(\textbf{h})h_k}{\theta P^\star_{\text{RAD}}(t)h_k + \sigma^2} \ge \Gamma, \quad \forall k \in \mathcal{K}_V.
\end{equation}
It is straightforward to obtain the optimum power allocation for each voice user as follows (realizing that at the optimum, constraints $C1$ are fulfilled with equality):
\begin{equation}
\check{p}^\star_k(\textbf{h}) = \frac{\Gamma(\theta P^\star_{\text{RAD}}(t)h_k + \sigma^2)}{M_Vh_k} , \quad \forall k \in \mathcal{K}_V.
\end{equation}
At this point, we could check the feasibility of \eqref{pro_1}. The problem is feasible if
\begin{equation}
T_s\sum_{k\in\mathcal{K}_V} \check{p}^\star_k(\textbf{h}) \le\phi (B(t)),
\end{equation}
which could also be written only in terms of the channels of the voice users, the current battery level, and some constants as follows:
 \begin{equation}
 \sum_{k\in\mathcal{K}_V} \frac{1}{h_k} \le \kappa_1 \phi(B(t)) - \kappa_2,
 \end{equation}
 where $\kappa_1 = \frac{M_V - |\mathcal{K}_V|\theta\Gamma}{\sigma^2T_s\Gamma}$ and $\kappa_2 = \frac{|\mathcal{K}_V|\theta P_{\text{CPICH}}}{\sigma^2}$. If the problem is not feasible, then we should consider either reducing such minimum SINR requirements (which will increase the constant term $\kappa_1$), or drop out some voice users in the scheduling period, or increase $\phi(B(t))$ by increasing the value of $\alpha$, but always guaranteeing that the maximum radiated constraint $P^{\max}_{\text{BS}}$ is not exceeded.


\subsection{Resource Allocation for Data Users}
Now that we have considered the voice users, we can tackle the resource allocation problem for the data users by solving problem \eqref{pro_1}. Note that problem \eqref{pro_1} is convex once we know $P^\star_{\text{BS}}(t)$. To solve problem, \eqref{pro_1}, we will reformulate it by introducing the slack variable $s$, which preserves convexity \cite{boyd}, as 
\begin{alignat}{2}
\mathop{\text{maximize}}_{s,\,\textbf{r},\, \textbf{p},\, \textbf{n}} & \quad s \label{pro_2} \\
\text{subject to} 
& \quad C2,\dots, C6 \text{\, of \,problem \,} \eqref{pro_1}\nonumber \\ 
& \quad C8: s \le \mathbb{E}_{\textbf{h}}[r_k(\textbf{h})], \quad \forall k \in \mathcal{K}_D \nonumber \\ 
& \quad C9: 0 \le s \le \frac{R_{BH}- \check{R}_{BH}(|\mathcal{K}_V|)}{\xi|\mathcal{K}_D|}. \nonumber
\end{alignat}

Notice that we have introduced an additional constraint, $C9$. As it is clear from the formulation, this constraint does not affect the optimum solution, but it will help in the numerical search of the optimum value of the new slack variable $s$. Notice also that the previous optimization problem is time-coupled (we require the future channel realizations due to the expectation operator appearing in $C8$). In order to deal with such difficult problem involving expectations, we propose to use a stochastic approximation that has been proposed in the literature \cite{ribeiro:10}. In this approach, the constraints involving expectations are dualized, and their Lagrange multipliers are estimated stochastically at each period. 

Let us start by dualizing constraint $C8$. Let $\boldsymbol\lambda \triangleq \{\lambda_k, \, \forall k \in\mathcal{K}_D\}$ be the vector of Lagrange multipliers associated to $C8$. The partial Lagrangian is given by $\mathcal{L}_{C8}(s,\boldsymbol\lambda) = -s + \sum_{k\in\mathcal{K}_D} \lambda_k\left(s-\mathbb{E}_{\textbf{h}}[r_k(\textbf{h})]\right)$. In order to find the optimum $s$ we have to perform the following minimization:
\begin{alignat}{2}
\mathop{\text{minimize}}_{0\le s \le \frac{R_{BH}- \check{R}_{BH}(|\mathcal{K}_V|)}{\xi|\mathcal{K}_D|}} & \quad \mathcal{L}_{C8}(s,\boldsymbol\lambda).
\label{par_lag}
\end{alignat}

According to author in \cite{ribeiro:10}, when the objective function is linear in the optimization variable, the stochastic primal-dual algorithms present some numerical problems. This can be avoided by transforming the objective function introducing a general differentiable monotonically increasing cost function $U(\cdot)$ (e.g., the logarithm). Note that the introduction of such function does not modify the optimal value of the optimization variables (i.e., the solution is the same). Given that, setting the gradient to zero, $\nabla_{s}\mathcal{L}_{C8}(s,\boldsymbol\lambda) = 0$ and solving yields:
\begin{equation}
s^\star(\boldsymbol\lambda) = \left((\dot{U})^{-1}\left(\sum_{k\in\mathcal{K}_D}\lambda_k\right)\right)^{\frac{R_{BH}- \check{R}_{BH}(|\mathcal{K}_V|)}{\xi|\mathcal{K}_D|}}_0,
\end{equation}
where $\dot{U}(\cdot)$ is the derivative of $U(\cdot)$ and $(\dot{U})^{-1}(\cdot)$ is the inverse function of $\dot{U}(\cdot)$. Once we know the optimum $s^\star$, the problem \eqref{pro_2} is updated as follows (where we have skipped in the objective function the term that does not depend on the optimization variables remaining in the optimization problem):
\begin{alignat}{2}
\mathop{\text{maximize}}_{\textbf{r},\, \textbf{p},\, \textbf{n}} & \quad \sum_{k\in\mathcal{K}_D}\lambda_k \mathbb{E}_{\textbf{h}}[r_k(\textbf{h})]  \\
\text{subject to} 
& \quad C2,\dots, C6 \text{\, of \,problem \,} \eqref{pro_1}.\nonumber
\end{alignat}
Now, we proceed to dualize constraint $C2$. Let $\boldsymbol\mu \triangleq \{\mu_k, \, \forall k \in\mathcal{K}_D\}$ be the vector of Lagrange multipliers associated to $C2$. The partial Lagrangian is
\begin{eqnarray}
\mathcal{L}_{C2}(r_k(\textbf{h}),\boldsymbol\lambda,\boldsymbol\mu) = &-&\sum_{k\in\mathcal{K}_D}\lambda_k \mathbb{E}_{\textbf{h}}[r_k(\textbf{h})] + \sum_{k\in\mathcal{K}_D} \mu_k\left(\mathbb{E}_{\textbf{h}}[r_k(\textbf{h})] -\frac{R_{BH}- \check{R}_{BH}(|\mathcal{K}_V|)}{\xi|\mathcal{K}_D|}\right),\\
= &-& \mathbb{E}_{\textbf{h}}\left[\sum_{k\in\mathcal{K}_D}(\lambda_k - \mu_k)r_k(\textbf{h})\right]  - \sum_{k\in\mathcal{K}_D} \mu_k\left(\frac{R_{BH}- \check{R}_{BH}(|\mathcal{K}_V|)}{\xi|\mathcal{K}_D|}\right).
\end{eqnarray}
For given Lagrange multipliers $\boldsymbol\lambda$ and $\boldsymbol\mu$, the optimization problem \eqref{pro_2} is equivalently reformulated as (where we have skipped again in the objective function the term that does not depend on the optimization variables remaining in the optimization problem):
\begin{alignat}{2}
\mathop{\text{maximize}}_{\textbf{r},\, \textbf{p},\, \textbf{n}} & \quad \sum_{k\in\mathcal{K}_D}(\lambda_k - \mu_k)r_k(\textbf{h}) \label{pro_3} \\
\text{subject to} 
& \quad C3,\dots, C6 \text{\, of \,problem \,} \eqref{pro_1}.\nonumber 
\end{alignat}
Notice that the expectations are no longer present in the formulation because the remaining constraints $C3-C6$ are applied to instantaneous resource allocation variables (without expectations) and also because the maximization of the expected value of the objective function with respect to $\textbf{r},\, \textbf{p}$, and $\textbf{n}$ in the current scheduling period, in this case, is the same as the maximization of the term within the expectation. The problem now resides in the computation of the optimum Lagrange multipliers which requires knowing the statistics of $r_k(\textbf{h})$. If we solve the dual problem of \eqref{pro_3}, i.e., $\sup_{\boldsymbol\lambda\succeq 0,\, \boldsymbol\mu \succeq 0} \inf \mathcal{L}(\textbf{r},\, \textbf{p},\, \textbf{n}, \boldsymbol\lambda,\boldsymbol\mu),$ where $\succeq$ means element-wise inequality and the Lagrangian, $\mathcal{L}$, is defined in Appendix \ref{app1}, using a gradient approach, then the optimum multipliers could be found as \cite{boyd}: 
\begin{eqnarray}
\lambda^{(q+1)}_k &=& \left(\lambda^{(q)}_k + \epsilon\left(s^\star(\boldsymbol\lambda^{(q)}) -\mathbb{E}_{\textbf{h}}\left[r^\star_k\left(\textbf{h}; \boldsymbol\lambda^{(q)}, \boldsymbol\mu^{(q)}\right)\right]\right)\right)^\infty_0, \quad \forall k,\label{opt_mul1}\\
\mu^{(q+1)}_k &=& \left(\mu^{(q)}_k + \epsilon\left(\mathbb{E}_{\textbf{h}}\left[r^\star_k\left(\textbf{h}; \boldsymbol\lambda^{(q)}, \boldsymbol\mu^{(q)}\right)\right] -  \frac{R_{BH}- \check{R}_{BH}(|\mathcal{K}_V|)}{\xi|\mathcal{K}_D|}\right)\right)^\infty_0, \,\, \forall k,\label{opt_mul2}
\end{eqnarray}
where $\epsilon$ is the step size. Note that, it is not possible to compute the value of the Lagrange multipliers in real time, and then solve \eqref{pro_3}, as they depend on the statistics of $r_k(\textbf{h})$ that is a function not known a priori (it is the solution of the optimization problem itself). In this situation, we propose to follow a stochastic approximation \cite{ribeiro:10} and eliminate such uncertainty constraint by estimating the multipliers stochastically at each scheduling period (with a noisy instantaneous unbiased estimate of the gradient) as follows (note that this philosophy is similar to the instantaneous estimation of the gradient in the LMS algorithm \cite{haykin_book}): 
\begin{eqnarray}
\lambda_k(t+1) &=& \left(\lambda_k(t) + \epsilon\left(s^\star(\boldsymbol\lambda(t)) - r^\star_k(\textbf{h}; \boldsymbol\lambda(t),\boldsymbol\mu(t))\right)\right)^\infty_0, \quad \forall k,\label{est_1}\\
\mu_k(t+1) &=& \left(\mu_k(t) + \epsilon\left(r^\star_k(\textbf{h}; \boldsymbol\lambda(t),\boldsymbol\mu(t)) - \frac{R_{BH}- \check{R}_{BH}(|\mathcal{K}_V|)}{\xi|\mathcal{K}_D|}\right)\right)^\infty_0, \quad \forall k. \label{est_2}
\end{eqnarray}

The advantages of the stochastic techniques are threefold: \emph{i}$)$ the computational complexity of the stochastic technique is significantly lower than that of their off-line counterparts; \emph{ii}$)$ stochastic approaches can deal with non-stationarity environments; \emph{iii}$)$ the distribution of the involved random variables $\textbf{h}$ is not required. 

Once we update the values of the Lagrange multipliers, problem \eqref{pro_3} can be solved using, for example a primal-dual approach. Notice that constraint $C3$ can be put directly in the objective function as, at the optimum, it is fulfilled with equality, i.e., $r^\star_k(\textbf{h}; \boldsymbol\lambda(t),\boldsymbol\mu(t)) = n^\star_k(\textbf{h}; \boldsymbol\lambda(t),\boldsymbol\mu(t))  \frac{W}{M_D}\log_2\left(1+\frac{M_Dp^\star_k(\textbf{h}; \boldsymbol\lambda(t),\boldsymbol\mu(t))h_k}{n^\star_k(\textbf{h}; \boldsymbol\lambda(t),\boldsymbol\mu(t))(\theta P^\star_{\text{RAD}}(t) h_k + \sigma^2)}\right)$. Thus, the resource allocation problem to be solved at the beginning of the scheduling period $t$ is given by
\begin{alignat}{2}
\mathop{\text{maximize}}_{\textbf{p},\, \textbf{n}} & \quad \sum_{k\in\mathcal{K}_D}(\lambda_k(t) - \mu_k(t))n_k  \frac{W}{M_D}\log_2\left(1+\frac{M_Dp_kh_k}{n_k(\theta P^\star_{\text{RAD}}(t) h_k + \sigma^2)}\right) \label{pro_4} \\
\text{subject to} 
& \quad C4,\dots, C6 \text{\, of \,problem \,} \eqref{pro_1}.\nonumber 
\end{alignat}
Notice that we have not considered the dependency of the optimization variables with respect to the channels explicitly for the sake of simplicity in the notation. Problem \eqref{pro_4} can be solved using, for example, a primal-dual approach, as it is given in Appendix \ref{app1}. 

It can be shown that the sample average of the stochastic rates, $r^\star_k(\textbf{h}; \boldsymbol\lambda(t), \boldsymbol\mu(t))$, satisfies all the constraints in \eqref{pro_1} and incurs minimal performance loss relative to the optimal (off-line) solution of \eqref{pro_1}. This can be stated rigorously as follows: define $F(t) \triangleq \min_{k\in\mathcal{K}_D} \frac{1}{t}\sum_{\tau=1}^t  r^\star_k(\textbf{h}; \boldsymbol\lambda(\tau), \boldsymbol\mu(\tau))$ and $f^\star$ as the minimum value of the objective function in \eqref{pro_1}. Then, it holds with probability one that as $t\rightarrow \infty$: \emph{i}$)$ the solution is feasible; and \emph{ii}$)$ $F(t) \le f^\star + \delta(\epsilon)$, where $\delta(\epsilon) \rightarrow 0$ as $\epsilon \rightarrow 0$. A proof of this result is not presented here due to space limitations but it can be derived following \cite{ribeiro:10}. Let us introduce some important remarks here regarding problem \eqref{pro_4}:

\textbf{\emph{Remark 1:}} the values of $\{\lambda_k\}$ measure how far the average rate $s^\star(\boldsymbol\lambda(t))$ is from the instantaneous rates served to the users in the access network. If the quality of the channels or the available powers are such that the instantaneous rates served in the access network are far from the target average rate $s^\star(\boldsymbol\lambda(t))$ for all the users, the sum of $\{\lambda_k\}$ will increase and the system will reduce the target average rate $s^\star(\boldsymbol\lambda(t))$. 

\textbf{\emph{Remark 2:}} in the access network, the values of $\{\lambda_k\}$ are in charge of ensuring that all average rates tend to grow simultaneously (maximin objective function). Note that, if at any temporal period, $s^\star(\boldsymbol\lambda(t)) > r^\star_k(\textbf{h}; \boldsymbol\lambda(t), \boldsymbol\mu(t))$, then $\lambda_k(t+1)$ grows and the priority of the $k$-th user to be served increases. 

\textbf{\emph{Remark 3:}} if at any temporal period, $r^\star_k(\textbf{h}; \boldsymbol\lambda(t), \boldsymbol\mu(t))>  \frac{R_{BH}- \check{R}_{BH}(|\mathcal{K}_V|)}{\xi|\mathcal{K}_D|}$, then $\mu_k(t+1) > \mu_k(t)$. For a fixed set of $\{\lambda_k\}$, if $\lambda_k(t+1)-\mu_k(t+1)$ decreases, the user will have a lower priority to be served in the next period. The same reasoning could be applied if $r^\star_k(\textbf{h}; \boldsymbol\lambda(t), \boldsymbol\mu(t))<  \frac{R_{BH}- \check{R}_{BH}(|\mathcal{K}_V|)}{\xi|\mathcal{K}_D|}$ to deal with the reverse situation.

\subsection{Resource Allocation Algorithm}
In this subsection, we just present the overall algorithm to solve the resource allocation for the voice and data users based on the approaches presented in previous sections. The stochastic updates, as well as the battery update, are also presented. The algorithm is summarized in Table \ref{alg1}. Notice that, in the algorithm, steps 7 to 15 correspond to the steps presented in Appendix \ref{app1} to solve the convex optimization problem \eqref{pro_4}.

\begin{table}[t]
\small
\renewcommand{\arraystretch}{1.2}
\caption{\textsc{Algorithm for Solving Resource Allocation Problem \eqref{pro_1}}}
\vspace{-0.5cm}
\label{alg1}
\begin{center}
\begin{tabular}{ l  l }
\hline\hline
1: & initialize $\lambda_k(t)\ge 0$, $\mu_k(t) \ge 0$, $\forall k\in\mathcal{K}_D$\\
2: & compute $P^\star_{\text{RAD}}(t) = \frac{\phi \left(B(t)\right)}{T_s} + P_{\text{CPICH}}$\\
3: & \underline{\textbf{Voice users}} \\
4: & \quad \quad compute $\check{p}^\star_k(\textbf{h}) = \frac{\Gamma(\theta P^\star_{\text{RAD}}(t)h_k + \sigma^2)}{M_Vh_k} , \quad \forall k \in \mathcal{K}_V$\\
5: & \quad \quad \textbf{if} \, $T_s\sum_{k\in\mathcal{K}_V} \check{p}^\star_k(\textbf{h}) > \phi (B(t))$ \,\, $\longrightarrow \,$  drop some voice users or reduce $\Gamma$, then go to 4\\
6: & \underline{\textbf{Data users}} \\
7: &\quad \quad\textbf{repeat}\\
8: & \quad \quad\quad \quad initialize $\textbf{n} \succeq 0$\\
9: & \quad \quad\quad \quad\textbf{repeat}\\
10: & \quad \quad\quad \quad \quad \quad $p_k^{(q,k+1)} = p^\star_k\left(\textbf{n}^{(q,k)}, \beta^{(q)}, \boldsymbol\lambda(t), \boldsymbol\mu(t)\right)$ using \eqref{opt_pow}, $\forall k\in\mathcal{K}_D$\\ 
11: & \quad \quad\quad \quad \quad \quad $n_k^{(q,k+1)} = n^\star_k\left(\textbf{n}^{(q,k)},\textbf{p}^{(q,k+1)}, \varphi^{(q)}, \boldsymbol\lambda(t), \boldsymbol\mu(t)\right)$ using fixed-point iteration in \eqref{opt_cod}, $\forall k\in\mathcal{K}_D$\\ 
12: & \quad \quad\quad \quad\textbf{until} $p_k^{(q,k+1)}$ and $n_k^{(q,k+1)}$ converge \\
13: &\quad \quad \quad \quad update the dual variables, $\beta^{(q+1)}$ and $\varphi^{(q+1)}$, using $p_k^{(q)}$ and $n_k^{(q)}$ with \eqref{lag1} and \eqref{lag2} \\
14: & \quad \quad\textbf{until} $\beta^{(q+1)}$ and $\varphi^{(q+1)}$ converge \\
15: & \quad \quad compute $r^\star_k(\textbf{h}; \boldsymbol\lambda(t), \boldsymbol\mu(t))$ with $p^\star_k(\textbf{n}, \beta)$ and $n^\star_k(\textbf{p}, \textbf{n}, \varphi)$\\
16: & \quad \quad update (dualized) primal variable:\\
17: & \quad \quad \quad \quad $s^\star(\boldsymbol\lambda(t)) = \left((\dot{U})^{-1}\left(\sum_{k\in\mathcal{K}_D}\lambda_k(t)\right)\right)^{ \frac{R_{BH}- \check{R}_{BH}(|\mathcal{K}_V|)}{\xi|\mathcal{K}_D|}}_0$\\
18: &\quad \quad update stochastic dual variables:\\
19: &\quad \quad \quad \quad $\lambda_k(t+1) = \left(\lambda(t) + \epsilon\left(s^\star(\boldsymbol\lambda(t)) - r^\star_k(\textbf{h}; \boldsymbol\lambda(t),\boldsymbol\mu(t))\right)\right)^\infty_0$\\
20: & \quad \quad\quad \quad$\mu_k(t+1) = \left(\mu(t) + \epsilon\left(r^\star_k(\textbf{h}; \boldsymbol\lambda(t),\boldsymbol\mu(t)) -   \frac{R_{BH}- \check{R}_{BH}(|\mathcal{K}_V|)}{\xi|\mathcal{K}_D|}\right)\right)^\infty_0$\\
21: & update battery with consumed energy and harvesting:\\
22: & \quad \quad $B(t+1) =  \left(B(t) - E(t) + H(t)\right)^{B^{\max}}_0$\\
23: & $t \longleftarrow t + 1$ and go to 2\\
\hline\hline
\end{tabular}
\end{center}
\vspace{-0.5cm}
\end{table}

\section{Numerical Evaluation}
\label{sec_num}
In this section we evaluate the performance of the proposed strategy. The scenario under consideration is composed of 1 BS, and 3 voice users and 6 data users. The maximum radiated power is $P^{\max}_{\text{BS}} = 9$ dBm, the pilot power is $P_{\text{CPICH}} = 4$ dBm (which represents the $13\%$ of the maximum radiated power, as we considered in \cite{tucand41}), and the fixed power is $P_c = 3$ dBm (considering the model in \cite{auer:11}, which was applied in \cite{tucand41}). The number of available codes for data transmission services is $N_{\max} = 15$. All the users are mobile with a speed of 3 m/s. The instantaneous channel gain, $h_k$, incorporates antenna gains, Rayleigh fading with unitary power, and a real path loss of a town in Per\'u known as San Juan (see details in \cite{tucand41}). The orthogonality factor is $\theta = 0.35$. The code gain of data codes $M_D = 16$ and the minimum SINR normalized with code gain for voice users is, $\frac{\Gamma}{M_V} = -13.7$ dB which corresponds to a rate of 12.2 Kbps. The noise power is $\sigma^2 = -102$ dBm. The battery capacity is $B^{\max} = 410$ $\mu$J, the energy packet size is $e = 30$ $\mu$J, and $\alpha = 0.3$ unless otherwise stated. The scheduling period for the data users and voice users are 2ms and 20ms, respectively, thus, $T_s = 2$ms. The utility function is $U(\cdot) = \log(\cdot)$. Two backhaul capacities have been considered in the simulations: $R_{BH} = 2$ Mbps and $R_{BH} = 500$ Kbps. The amount of backhaul capacity required by the 3 voice users considered in this deployment is $\check{R}_{BH}(|\mathcal{K}_V|) = 173$ Kbps. The overhead for the data transmissions is $\xi = 1.2$. The step size for the update of the stochastic multipliers is $\epsilon = 10^{-3}$. For a more detailed description of the simulation parameters see \cite{tucand41}. 

In the simulations, we consider as a benchmark the case where the BS is connected to the electric grid (which means equivalently that the battery remains full of energy for the whole simulation). For comparison purposes, we also show the resource allocation of the proportional fair (PF) strategy \cite{wang:07} with an instantaneous per-user backhaul constraint, $r_k(\textbf{h}) \le  \frac{R_{BH}- \check{R}_{BH}(|\mathcal{K}_V|)}{\xi|\mathcal{K}_D|}$, and an instantaneous sum constraint, $\sum_{k\in\mathcal{K}_D}r_k(\textbf{h}) \le  \frac{R_{BH}- \check{R}_{BH}(|\mathcal{K}_V|)}{\xi}$. Note that our proposed stochastic scheduling strategy resulting from \eqref{pro_1} cannot be compared directly in a fair way with the PF approach under a sum-rate constraint. The reason is that in \eqref{pro_1} we divide equally the backhaul capacity among users in average terms whereas in the PF with sum-rate constraint, the overall backhaul capacity is not forced to be distributed equally. The effective length of the exponential window in the PF has been set to $T_c = 500$ \cite{wang:07}\footnote{The weights of the PF scheduler are calculated as $\omega_k(t) = \frac{1}{T_k(t)}$, where $T_k(t)$ is the average throughput of user $k$ computed as $T_k(t) =  (1-\frac{1}{T_c})T_k(t-1) + \frac{1}{T_c}r_k(t)$.}.

\begin{figure}[t]
\centering
\includegraphics[width = 0.9\textwidth]{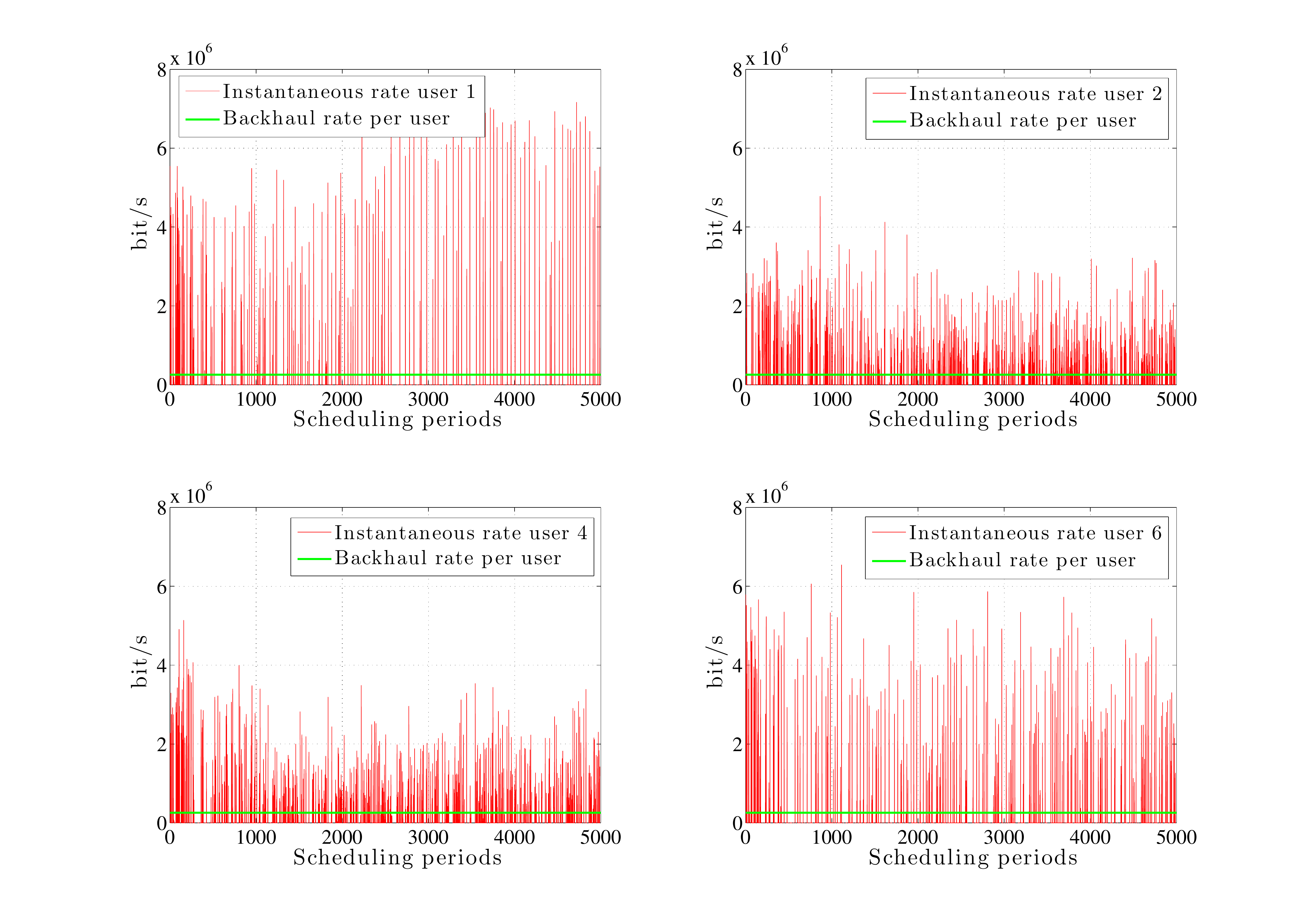}
\vspace{-2mm}
\caption{Time evolution of the instantaneous data rates served at the access network and the backhaul capacity limitation per user with a backhaul capacity of $2$ Mbps.}
\label{fig1}
\vspace{-3mm}
\end{figure}

\begin{figure}[t]
\centering
\includegraphics[width = 0.8\textwidth]{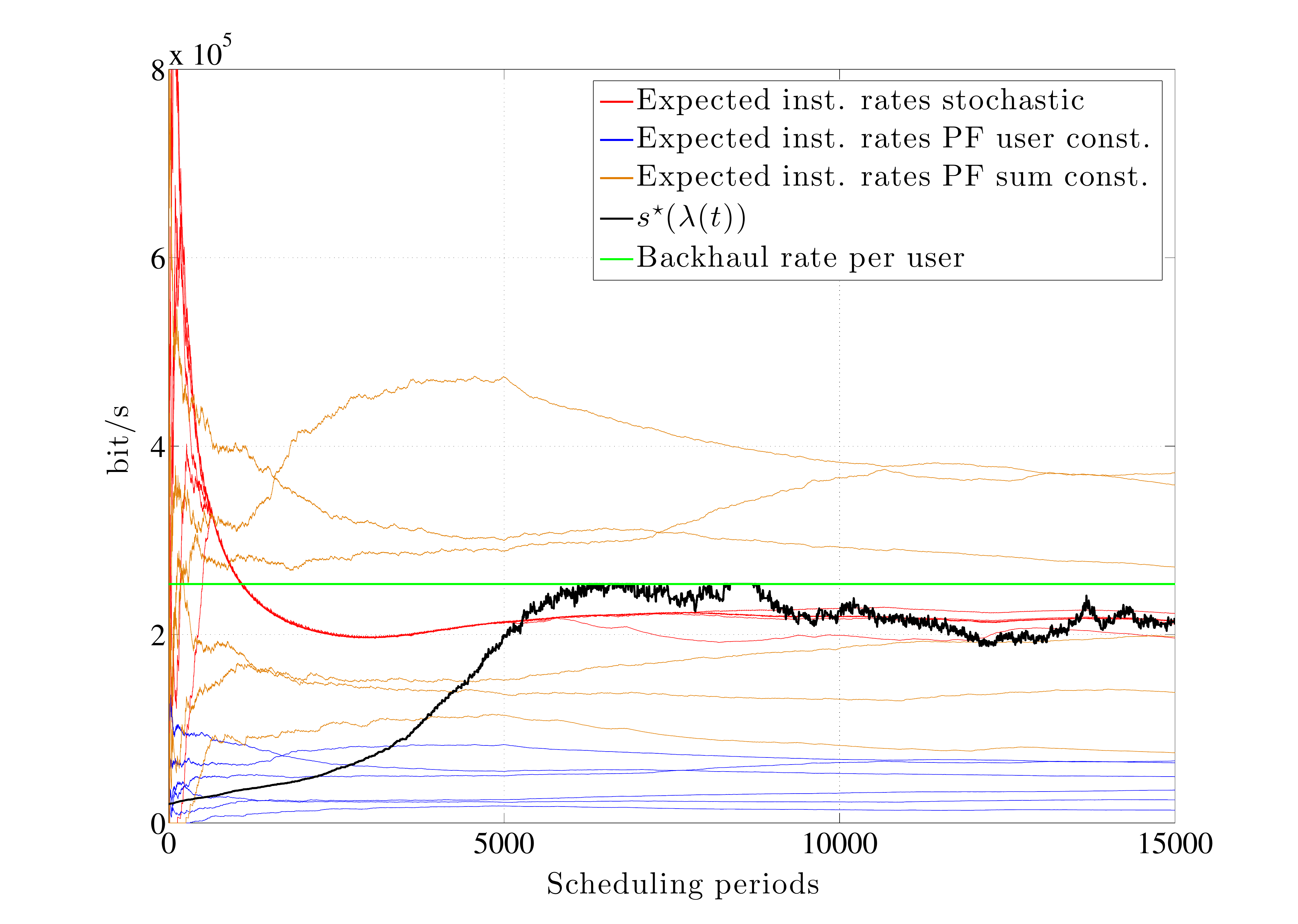}
\vspace{-2mm}
\caption{Time evolution of the data rates for the different approaches and the backhaul capacity per user when BS is connected to grid with a backhaul capacity of $2$ Mbps.}
\label{fig2}
\vspace{-2mm}
\end{figure}

Fig. \ref{fig1} presents the instantaneous data rates served at the access network of four data users out of the six. In this case, the BS is connected to the electric power grid. As we can see, the instantaneous rates are able to exceed the backhaul capacity in particular scheduling periods whereas, at the same time, the average rates fulfills the maximum backhaul capacity as it is shown in Fig. \ref{fig2}.

\begin{figure}[t]
\centering
\includegraphics[width = 0.8\textwidth]{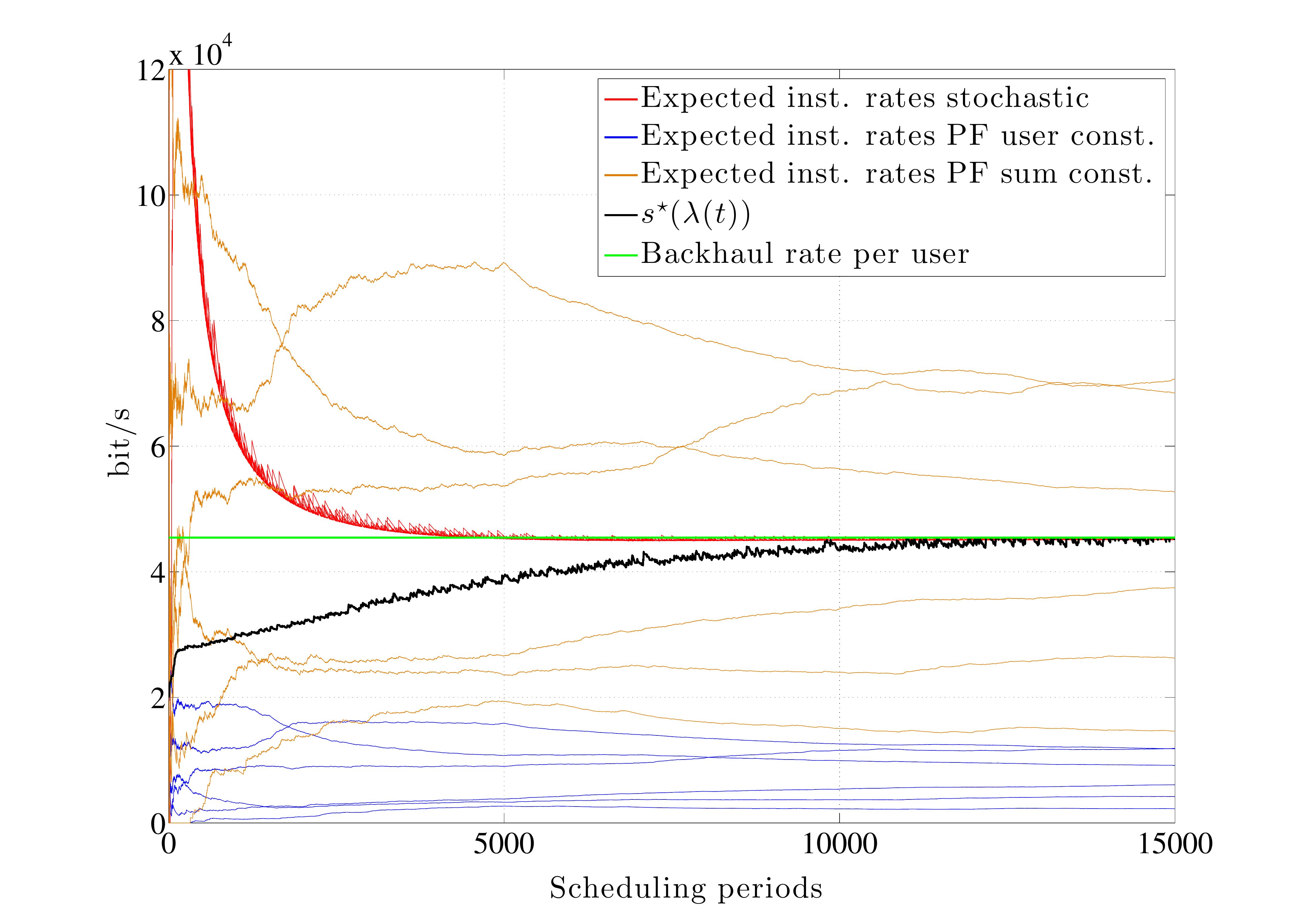}
\vspace{-2mm}
\caption{Time evolution of the data rates for the different approaches and the backhaul capacity per user when BS is connected to grid with a backhaul capacity of $500$ Kbps.}
\label{fig3}
\vspace{-2mm}
\end{figure}

Fig. \ref{fig2} shows the time evolution of the expected data rates of the three approaches. At any time instant the expected rates have been estimated using $r_k(t) = \frac{1}{t}\sum_{\tau=1}^{t} r_k(\tau)$. We also plot the time evolution of $s^\star(\boldsymbol\lambda(t))$ and the per-data user backhaul rate. In this case, the BS is connected to the electric grid. The backhaul capacity is $R_{BH} = 2$ Mbps. Initially, we assume that the queues at the access network are sufficiently full so that all the bits demanded by the users are served. This makes the initial average rates violate the backhaul capacity constraint for a short period of time (see the initial transient in the figure). This is due to the stochastic approximation of the multipliers but, in any case, when the average rates converge, they fulfill all the constraints of the original problem. As we can also see from the figure, the limitation of the rates comes from the limited resources available at the access network, i.e., the power and the codes, as the backhaul capacity is not reached. It should be also emphasized that, the proposed stochastic approach provides a solution that introduces more fairness when compared with the PF approach as the average rates for the different users are quite similar. Fig. \ref{fig3} depicts the same curves but now considering a backhaul capacity of $R_{BH} = 500$ Kbps. As we can see, now the system is limited by the backhaul and not by the limited resources of the access network.

\begin{figure}[t]
\centering
\includegraphics[width = 0.8\textwidth]{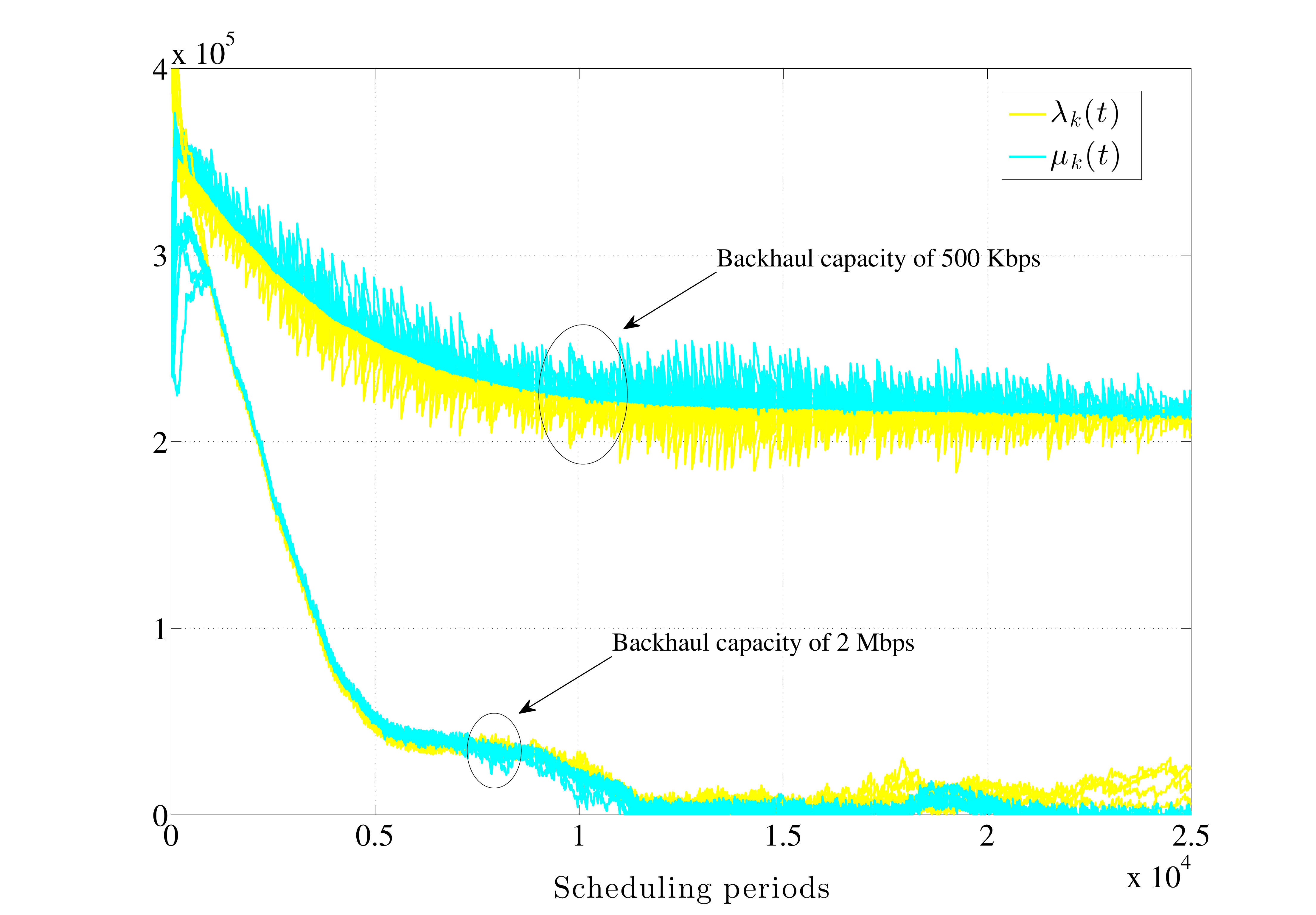}
\vspace{-2mm}
\caption{Time evolution of the stochastic Lagrange multipliers for different backhaul capacities.}
\label{fig4}
\vspace{-2mm}
\end{figure}

Fig. \ref{fig4} shows the evolution of the stochastic estimation of the Lagrange multipliers, i.e., $\lambda_k(t)$ and $\mu_k(t)$, for the cases where we have a backhaul capacity of $R_{BH} = 2$ Mbps and the case of having a backhaul capacity of $R_{BH} = 500$ Kbps. From duality theory, we know that if the backhaul constraint is not active, i.e., if the expected rates are below the backhaul capacity, then the optimum value of the multipliers is zero. This is what we see in the figure for the case of having a backhaul capacity of 2 Mbps. On the other hand, if the system is limited by the backhaul, then the optimum Lagrange multipliers are generally not zero as the corresponding constraints become active. From the figure we see that the multipliers converge to a non-zero value. The convergence of the multipliers states that the overall stochastic approach is working properly.

\begin{figure}[t]
\centering
\includegraphics[width = 0.8\textwidth]{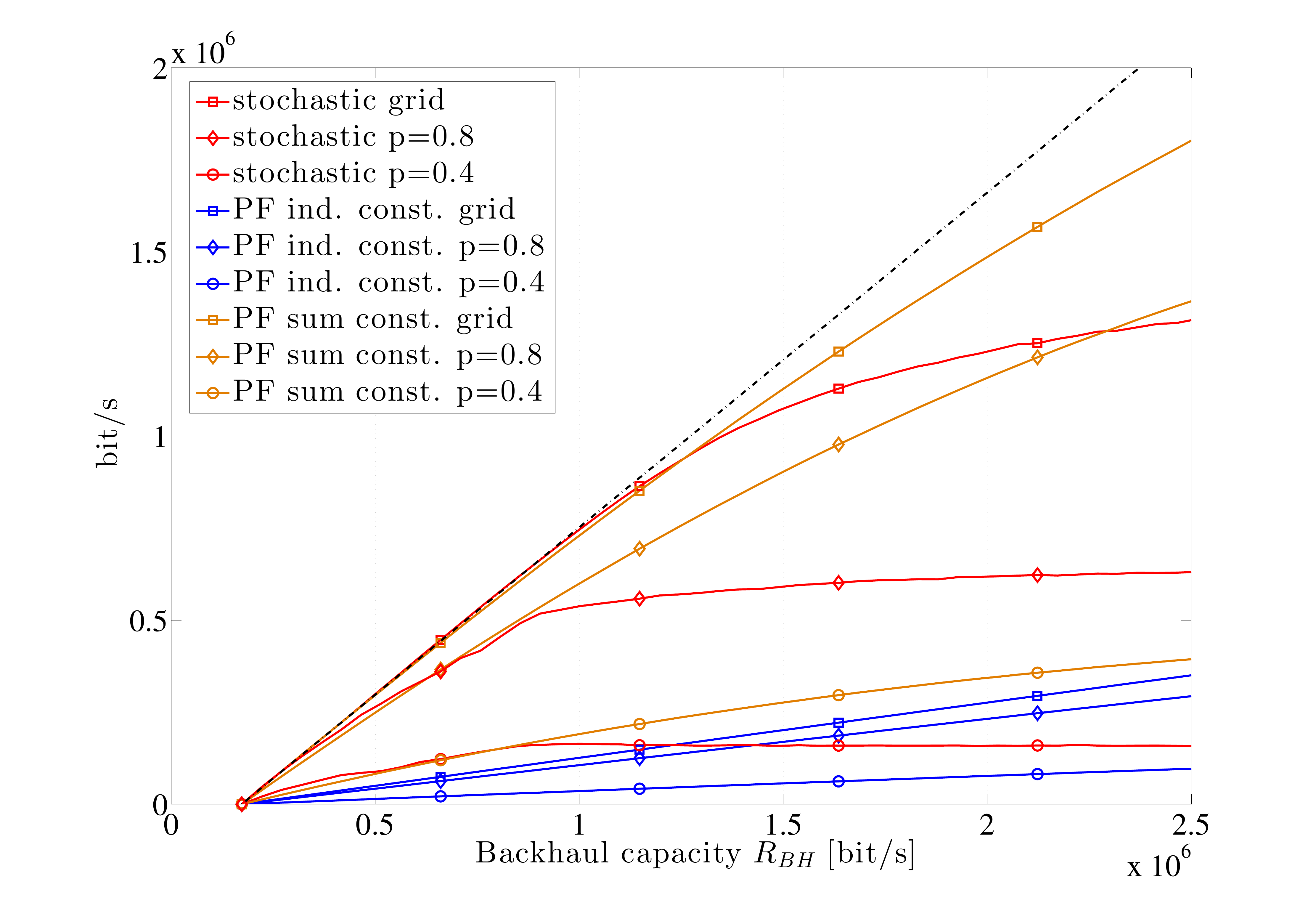}
\vspace{-2mm}
\caption{Sum rates as a function of backhaul capacity for different approaches and different probability of energy packet $p$.}
\label{fig5}
\vspace{-3mm}
\end{figure}

Fig. \ref{fig5} depicts the sum of the average data rates $\left(\frac{1}{t}\sum_{\tau=1}^{t}\sum_{k\in\mathcal{K}_D}r_k(\tau)\right)$ as a function of the overall backhaul capacity $(R_{BH})$ for the different approaches when the BS is connected to the grid and when the BS has a finite battery with different harvesting intensities $p$. The black dashed line shows the total available backhaul for data users. In this figure, we are able to identify when the system is limited by backhaul and when the system is limited by the access network. For example, for the stochastic case connected to the grid, the system is limited by the access network when the capacity of the backhaul is above $1.2$ Mbps. Concerning the comparison between our strategy and the PF approaches, we would like to emphasize again that the comparison with the PF approach under sum-rate constraint is not fair since in the last case the rate in the backhaul is not forced to be equally divided among users. Note, however, that even in that case, the sum-rate achieved by the PF is not much higher than the one achieved by our proposed stochastic approach. In addition, our proposed solution provides much more fairness as shown in Fig. \ref{fig2} and Fig. \ref{fig3}.

\begin{figure}[t]
\centering
\includegraphics[width = 0.8\textwidth]{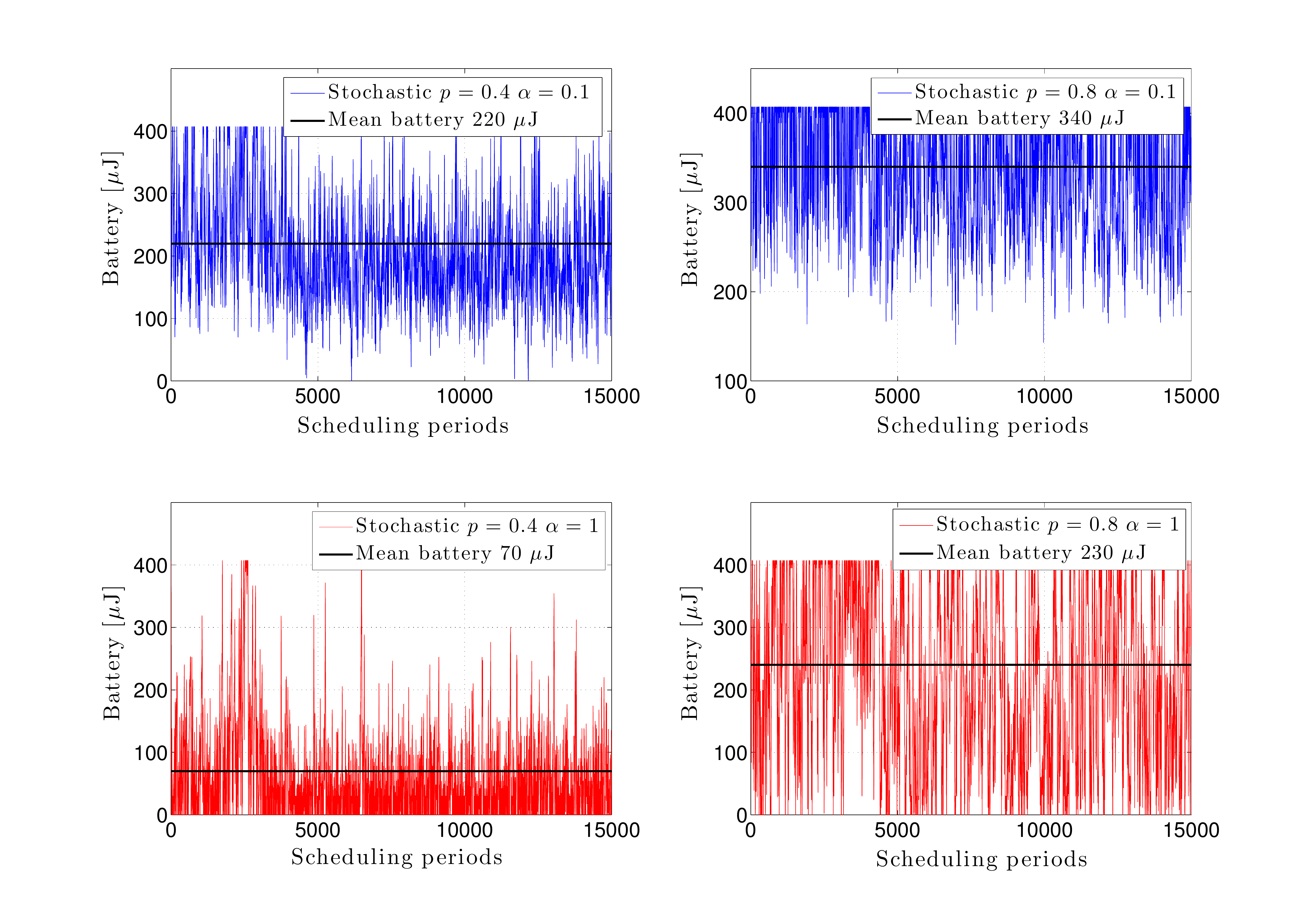}
\vspace{-2mm}
\caption{Battery evolution of the proposed stochastic approach and the PF with sum constraint with a probability of energy packet $p=0.4$ and $p=0.8$ a for $\alpha=0.1$ and $\alpha = 1$.}
\label{fig6}
\vspace{-3mm}
\end{figure}

Finally, Fig. \ref{fig6} depicts the time evolution of the instantaneous battery level of the BS when the electric grid is not available for the stochastic approach. We assume that the probability of receiving an energy packet during one scheduling period is $p=0.4$ and $p=0.8$ and two values of $\alpha$ have been considered: $\alpha=0.1$ and $\alpha=1$. Recall that $\alpha=1$ means that all battery could be used during one particular scheduling period (if the physical limitation of the BS allows it). In can be proved that, if no physical power limitation exists at the BS and the battery never reaches its maximum value, then, theoretically the expected value of the battery is given by $\hat{b} = \frac{\mathbb{E}[H(t)]}{\alpha} = \frac{p\cdot e}{\alpha}$. However, because of the maximum power radiation at the BS and the battery overflows, the previous expression yields a lower bound of the true expected battery level, i.e., $\hat{b} \le \lim_{t\rightarrow\infty}\mathbb{E}[B(t)]$. For example, if $p=0.8$, $\alpha = 0.1$, then $\hat{b}= 240$ $\mu$J, but the figures shows $\lim_{t\rightarrow\infty}\mathbb{E}[B(t)] = 340$ $\mu$J. The larger the value of $\alpha$, the looser the lower bound is.

\vspace{-3mm}
\section{Conclusions}
\label{sec_con}
In this paper, we have proposed a resource allocation strategy based on the maximization of the minimum average data rates in a WCDMA system. By the use of stochastic optimization tools, we are able to consider a backhaul capacity constraint in terms of the average data rate, allowing the access network to offer higher rates taken advantage of good instantaneous wireless channel conditions. We have assumed that the BS is powered with a finite battery that is able to be recharged by means of a harvesting source. The dynamics of the energy harvesting, the energy spending, and the battery have also been taken into account explicitly in the proposed resource allocation problem. Simulations results showed that the proposed approach achieves more fairness among the users when compared to the traditional PF strategy and, for some backhaul capacities, the sum-rate is higher if compared with the PF with an instantaneous per-user backhaul constraint, or the same if compared with the PF with an instantaneous sum-rate backhaul constraint. 

\appendices
\section{}
\label{app1}
In this appendix, we present the technique to solve \eqref{pro_4} assuming that the Lagrange multipliers are known (therefore, we omit the explicit dependence of the optimization variables with respect to the stochastic Lagrange multipliers, $\boldsymbol\lambda(t), \,\boldsymbol\mu(t)$, due to simplicity in the notation). Let $\beta$ and $\varphi$ be the Lagrange multipliers associated to constraints $C4$ and $C5$. There is no need to dualize constraint $C6$ because the solution will turn out to automatically satisfy it. The Lagrangian of problem \eqref{pro_4} is given by 
\begin{eqnarray}
\mathcal{L}\left(\textbf{p}, \textbf{n},\beta, \varphi \right) = &-& \sum_{k\in\mathcal{K}_D}(\lambda_k(t) - \mu_k(t))n_k \frac{W}{M_D}\log_2\left(1+\frac{M_Dp_kh_k}{n_k(\theta P^\star_{\text{RAD}}(t) h_k + \sigma^2)}\right) \\
&+& \beta \left(\sum_{k\in\mathcal{K}_D} p_k - \left(\frac{\phi(B(t))}{T_s} - \sum_{k\in\mathcal{K}_V}\check{p}_k\right)\right) + \varphi\left( \sum_{k\in\mathcal{K}_D} n_k - N_{\max}\right).\nonumber 
\end{eqnarray}
For given Lagrange multipliers, $\beta$ and $\varphi$, we need to minimize the Lagrangian w.r.t. the primal variables. As it will be shown next, the structure of $\mathcal{L}\left(\textbf{p}, \textbf{n},\beta, \varphi \right)$ allows the minimization w.r.t. $\textbf{p}$ and $\textbf{n}$ to be found in closed-form. Because $\mathcal{L}\left(\textbf{p}, \textbf{n},\beta, \varphi \right)$ is strictly convex and differentiable w.r.t. $\textbf{p}$ and $\textbf{n}$, minimization w.r.t. these variables requires to equating the corresponding partial derivatives of $\mathcal{L}\left(\textbf{p}, \textbf{n},\beta, \varphi \right)$ to zero. Differentiating the Lagrangian w.r.t. the data powers, equating the derivative to zero and solving such expression for the data powers yields
\begin{equation}
p^\star_k(\textbf{n}, \beta) = \left(\frac{(\lambda_k(t)-\mu_k(t))n_kW}{\ln(2)\beta M_D} - \frac{n_k\left(\theta P^\star_{\text{RAD}}(t) h_k + \sigma^2\right)}{M_Dh_k}\right)_0^\infty,
\label{opt_pow}
\end{equation}
where the projection on the nonnegative orthant guarantees that constraint $C6$ is fulfilled. Proceeding similar with the optimum code allocation, we set the partial derivative of the Lagrangian w.r.t. $n_k$, equate it to zero and solve such equation for the codes, yielding:
\begin{equation}
n^\star_k(\textbf{p}, \textbf{n}, \varphi) = \left(\frac{(\lambda_k(t)-\mu_k(t))WM_Dp_kh_k\left((\theta P^\star_{\text{RAD}}(t) h_k + \sigma^2)\ln(2)\right)^{-1}}{(\lambda_k(t)-\mu_k(t))W\log\left(1 + \frac{M_Dp_kh_k}{n_k(\theta P^\star_{\text{RAD}}(t) h_k + \sigma^2)}\right) - M_D\varphi} - \frac{M_Dp_kh_k}{\theta P^\star_{\text{RAD}}(t) h_k + \sigma^2}\right)_0^\infty,
\label{opt_cod}
\end{equation}
where, also in this case, the projection on the nonnegative orthant guarantees that constraint $C6$ is fulfilled. Notice that a fixed-point iteration to compute the optimum code allocation, $n^\star_k(\varphi)$, can be used in this case. 

Having obtained the optimum primal variables as a function of the Lagrange multipliers, we now seek to find the optimum Lagrange multipliers to obtain the global optimum primal variables. The approach we propose to find the optimum multipliers is based on the maximization of the dual function $D(\beta, \varphi)$ \cite{boyd}, which is defined as the minimization of the Lagrangian w.r.t. the primal variables, i.e., $D(\beta, \varphi) \triangleq \inf_{\textbf{p}, \textbf{n}} \mathcal{L}\left(\textbf{p}, \textbf{n},\beta, \varphi \right) \equiv \mathcal{L}\left(\{p^\star_k(\textbf{n}, \beta)\}, \{n^\star_k(\textbf{p}, \textbf{n}, \varphi)\}, \beta, \varphi\right)$. Then, the the multipliers are obtained by solving the dual problem as
\begin{alignat}{2}
\mathop{\text{maximize}}_{\beta, \varphi} &\quad D(\beta, \varphi)\\
\text{subject to} 
& \quad \beta \ge 0, \quad \varphi \ge 0.\nonumber 
\end{alignat}

Recall that the dual problem is always convex w.r.t. the dual variables and, thus, can be efficiently solved with a projected gradient method (if $D(\beta, \varphi)$ is differentiable) or a projected supergradient method if it is not differentiable \cite{bertsekas}. A valid supergradient for each particular dual variable is given by the constraint it is associated with \cite{bertsekas}. The update equations are given by
\begin{eqnarray}
\beta^{(q+1)} &=& \left(\beta^{(q)} + \nu^{(q)}\left(\sum_{k\in\mathcal{K}_D} p^{(q)}_k - \left(\frac{\phi(B(t))}{T_s} - \sum_{k\in\mathcal{K}_V}\check{p}_k\right)\right) \right)_0^\infty,\label{lag1}\\
\varphi^{(q+1)} &=& \left(\varphi^{(q)} +  \nu^{(q)}\left( \sum_{k\in\mathcal{K}_D} n^{(q)}_k - N_{\max}\right)\right)_0^\infty, \label{lag2}
\end{eqnarray}
where $q$ indicates the iteration and the step size defined as $\nu^{(q)} = \frac{Q}{\sqrt q}\left(\|\nabla D\|_2\right)^{-1}$, being $\nabla D$ the overall supergradient of the dual function, is chosen such that the diminishing conditions are fulfilled, i.e., $\lim_{k\rightarrow \infty} \nu^{(q)} = 0$, $\sum_{k\in\mathcal{K}_D}  \nu^{(q)} = \infty$ \cite{bertsekas}. Once we know the optimal dual variables, $\beta^\star$ and $\varphi^\star$, we can obtain the optimum power and code allocations, $p^\star_k(\beta^\star)$ and $n^\star_k(\varphi^\star)$. The proposed iterative algorithm is based on the primal-dual block coordinate descent method for the update of the primal variables $p_k$ and $n_k$ \cite{bertsekas}.

\bibliography{referencias}
\bibliographystyle{ieeetr}

\newpage

\end{document}